\documentclass[a4paper,12pt,pre,preprint]{revtex4}
\usepackage{amsmath}
\usepackage{amsthm}
\usepackage{amssymb}
\usepackage{graphicx}

\newcommand{\med}[1]{\widetilde{#1}}
\newcommand{\mean}[1]{\left\langle #1 \right\rangle}
\newcommand{\var}[1]{\mathrm{var}\!\left( #1 \right)}
\newcommand{\la}{\mathcal{L}_1}
\newcommand{\lb}{\mathcal{L}_2}

\newcommand{\dd}{\mathrm{d}}

\begin{document}

\title{Complete graph asymptotics for the Ising and random cluster
  models on 5D grids with cyclic boundary}

\date{\today} 

\author{P. H. Lundow} 
\email{per.hakan.lundow@math.umu.se} 

\author{K. Markstr\"om}
\email{klas.markstrom@math.umu.se} 

\affiliation{ Department of mathematics and mathematical statistics,
  Ume\aa{} University, SE-901 87 Ume\aa, Sweden}

\begin{abstract}
  The finite size scaling behaviour for the Ising model in five
  dimensions, with either free or cyclic boundary, has been the
  subject for a long running debate. The older papers have been based
  on ideas from e.g. field theory or renormalization.  In this paper
  we propose a detailed and exact scaling picture for critical region
  of the model with cyclic boundary.  Unlike the previous papers our
  approach is based on a comparison with the existing exact and
  rigorous results for the FK-random-cluster model on a complete
  graph.  Based on those results we identify several distinct scaling
  regions in an $L$-dependent window around the critical point.  We test
  these predictions by comparing with data from Monte Carlo
  simulations and find a good agreement.  The main feature which
  differs between the complete graph and the five dimensional
  model with free boundary is the existence of a bimodal energy
  distribution near the critical point in the latter.  This feature was
  found by the same authors in an earlier paper in the form of a
  quasi-first order phase transition for the same Ising model.
\end{abstract}


\maketitle

\section{Introduction}
The Ising model is one of the most studied models in statistical
physics. We now have a well developed mathematical theory for it's
behaviour in 2-dimensions \cite{SM1,SM2,grimmett2004random}, extensive
numerical work in 3-dimensions, and it is known that the upper
critical dimension for the model if $d=4$. For $d\geq 4$ it is known
that the model takes on it's mean field critical exponents in the
thermodynamic limit.  For finite systems far less is known rigorously
and there has been a long running debate on differences in scaling for
systems with free and cyclic boundaries
\cite{O1,O2,O3,O4,O5,O6,O7,O8,O9,O10,O11,O12}.  In a previous paper
\cite{boundarypaper} we found an hitherto overlooked way in which for
$d=5$ the behaviour of the model depends strongly on the boundary
condition, in fact so strongly that at the critical point the
difference is not believed to vanish as the side $L$ of the system
grows.  This conclusion has been questioned \cite{berche:08} by a
group of authors which has earlier been promoting an alternative
scaling picture for the case with free boundary conditions. However,
in a recent paper we revisited this question with much larger system
sizes and found a good agreement with our earlier results, and the
standard scaling picture, where the susceptibility of the model with
free boundary grows as $L^2$ and for cyclic boundary grows as
$L^{5/2}$.

In view of these results a natural question is why we should see a
different behaviour for the case with cyclic boundary, and what the
exact form of the scaling behaviour for this case should be for $d$
above the critical dimension. Our approach to the answer of this
question is via the Random-cluster model.  The Fortuin-Kasteleyn
random cluster model, or RC-model for short, is a natural extension of
the Ising, Potts and edge percolation models, all captured by varying
$q$, one of its two parameters, the other $p$ corresponding to the
temperature in the first two models and the edge probability in the
last.  Many properties of the Ising model have direct interpretations
in terms of the numbers and sizes of connected clusters in the
RC-model. In particular the susceptibility corresponds to the average
cluster size.

For percolation, the case $q=1$ of the RC-model, Aizenman
\cite{aizenman:97} conjectured that the largest clusters should scale
as $L^{2d/3}$ for $d>6$ with cyclic boundary, instead of $L^4$ for
free boundary. One of the reasons for this conjecture was that this
gives the same scaling as for the largest connected in the
Erd\"os-Renyi, or ER for short, random graph with $N$ vertices at it's
critical point, where the largest component has size proportional to
$N^{2/3}$.  Note that the ER-random graph can be seen as the
percolation model on the complete graph on $N$ vertices. This
conjecture was proved in \cite{HH:07,HH:11}, and in
\cite{MR2155704,MR2165583,MR2260845} it was proved that asymptotics of
the same type as on the complete graph can be expected on a wider
class of finite graphs as well.

As we have pointed out percolation is a special case of the RC-model
and the latter has also been studied on the complete graph by
mathematicians.  In \cite{MR1376340} the critical probability was
identified, the exponential asymptotics of the partition function was
studied, and it was proved that the phase transition is of second
order for $0\leq q \leq 2$ and of first order for $q>2$. Later
\cite{luczak:06} a more detailed study of the cluster structure was
carried out, and it was found that there are three ranges of $q$ with
distinct behaviour $q<2$, $q>2$ and $q=2$.

Our aim in this paper is to compare the behaviour of the largest and
second largest cluster in the RC-model for $q=2$, corresponding to the
Ising model, for 5-dimensional lattices with side $L$ and cyclic
boundary.  For the case $q=2 $ \cite{luczak:06} identified no less
than five distinct scaling regions near the critical probability for
the complete graph, each with its own asymptotic behaviour for $\la$
and $\lb$, the sizes of the largest and second largest clusters
respectively. Some of the regions are difficult to study, since in
order to obtain correct scaling they would require far larger graphs
than those used here.  Thus we have focused on three cases: i) the
high-temperature case (fixed $K$ for $K<K_c$), ii) near $K_c$ at fixed
$\lambda< 0$, and iii) near $K_c$ at fixed $\kappa$, where $\kappa$
and $\lambda$ are different $L$-dependent couplings.  As we shall
see, for these regions we have an excellent agreement between the
scaling for the complete graph and for the 5-dimensional Ising model
with cyclic boundary.

\section{Terminology, definitions and sampling details}
For a graph $G=G(V,E)$ on
$n=|V|$ vertices and $m=|E|$ edges the random cluster model's partition function is
\begin{equation}\label{ZRC}
  Z_{\mathrm{RC}}(G; p, q) = \sum_{A\subseteq E} p^{|A|}\,(1-p)^{|E|-|A|} q^{c(A)}
\end{equation}
where $c(A)$ is the number of (connected) components, or clusters as
we will call them, of the graph $G(V,A)$, i.e. the graph with vertex
set $V$ and edge set $A$. Note that $0<p<1$ and $q>0$ are parameters
to the distribution. The Ising model without an external field, on the
other hand, has the partition function
\begin{equation}\label{ZI}
  Z_{\mathrm{I}}(G; K) = \sum_{s\in\{\pm 1\}^V} \exp(K U(s))
\end{equation}
where the sum is taken over all functions $s$ from $V$ to $\pm
1$. Thus vertex $i$ has spin $s_i$.  Here the energy is
$U(s)=\sum_{ij\in E} s_i s_j$, summed over all edges in $G$.  The
parameter $K=1/T$ is the dimensionless coupling, or inverse
temperature. 

The two models are actually equivalent for $q=2$ when setting
$p=1-\exp(-2K)$. Assuming we have a state $A\subseteq E$ from the
random cluster distribution in Equation~\eqref{ZRC}, we can obtain a
state $s$ from the Ising distribution~\eqref{ZI} in the following way:
for each component, pick a spin $\pm 1$ uniformly at random and assign
this spin to all vertices of the component. Going in the other
direction is also easy.  Starting with a state $s$ from the Ising
distribution at coupling $K$, let $A=\emptyset$.  Now add each
satisfied edge, i.e. edges $ij$ with $s_is_j=1$, to the set $A$ with
probability $p=1-\exp(-2K)$.  For more information on this see
\cite{grimmett2004random}.

This second scheme is a surprisingly efficient way of obtaining random
cluster data. Just start up your trusted Ising state generator,
whether it be Metropolis, heat-bath or Wolff cluster
\cite{newman-barkema}, and convert the Ising states to correctly
distributed random cluster states.

All graphs studied here are 5-dimensional grid graphs with periodic
boundary conditions, i.e. cartesian products of five cycles on $L$
vertices, so that $n=L^5$ and $m=5L^5$, and we have used $L=4$, 6, 8,
10, 12, 16, 20 and 24. We collected the data by generating Ising
states with the Wolff cluster method~\cite{wolff:89} and then
converting them to random cluster states using the scheme described
above.

Throughout we use the critical coupling $K_c=0.113915$
\cite{boundarypaper}, thus corresponding to the random cluster
critical probablity $p_c\approx 0.203740$, and denote one scaled
temperature by $\kappa=n^{1/2}(K-K_c)/K_c$ and another by
$\lambda=n^{1/3}(K-K_c)/K_c$.  Recall that the critical temperature
for a complete graph on $n$ vertices approaches zero as $K_c\sim 1/n$
and hence $p_c\sim 2/n$.  We will denote a scaled probability by
$\varepsilon = (p-p_c)/p_c$.

The $k$th central moment of a distribution is denoted by $\sigma_k$,
the mean value by $\mean\cdots$ and the variance by $\var\cdots =
\sigma_2$. The standard deviation is then
$\sigma=\sqrt{\sigma_2}$. The median is written $\med\cdots$. Given a
random cluster state $A$ the largest and second largest cluster size,
i.e. the number of vertices in these clusters, is denoted $\la$ and
$\lb$ respectively.

\subsection{Scaling for the complete graph}
We will here give a very brief description of some of the relevant
scaling results from \cite{luczak:06} for the complete graph on $n$
vertices. Where relevant further details will be given in later
sections.  As proven in \cite{MR1376340} the critical probability is
given by $p_c=2/n$.  Now if $p/p_c \rightarrow a\neq 1$ then the model
is not critical and depending on whether $a$ is less than 1 or greater
than 1 we see behaviour corresponding to the high- and low-temperature
regions respectively, with only small clusters in the first case and a
large cluster, linear in $n$ sized, plus some small ones in the latter
case.

If $p/p_c \rightarrow 1$ we are inside the critical window and need a
finer parameterization of $p$. We thus assume that $n/2 p= 1
+\epsilon$, where $\epsilon$ can depend on $n$. We now see five
distinct regions inside the critical window:
\begin{enumerate}
\item if $n^{1/3}\epsilon \rightarrow -\infty$ then, asymptotically,
  all clusters are trees and small.
\item if $n^{1/3}\epsilon \rightarrow c<1$, where $c$ is a
constant, then $\la$ is roughly of order $n^{2/3}$. 
\item if
$n^{1/3}\epsilon \rightarrow 0$ but $n^{1/2}\epsilon \rightarrow
-\infty$ there exists a unique largest component and it's size is
$n^{c}$, for a value $2/3<c<3/4$.  
\item if $n^{1/2}\epsilon \rightarrow c$, where $c$ is a constant,
  then $\la$ is of order $n^{3/4}$ and $\lb$ is bounded by
  $\mathcal{O}(\log{n}\sqrt{n})$.
\item if $n^{1/2}\epsilon \rightarrow \infty$ and $\epsilon=o(1)$ then
  $\la$ is of order $n\sqrt{3\epsilon}$ and $\lb$ of order
  $\frac{\log{n^2\epsilon^{3}}}{\epsilon}$.
\end{enumerate}
 
\section{The high-temperature region}

Strictly speaking, the complete graph version of the high-temperature
case only requires that $\varepsilon\,n^{1/3}\to -\infty$. Of course,
a fixed $\varepsilon<0$ will satisfy this condition. Hence, for the 5D
case we will simply test the case of a fixed $K$ for $K<K_c$.

\begin{figure}[!hbt]
  \begin{minipage}{0.49\textwidth}
    \begin{center}
      \includegraphics[width=0.99\textwidth]{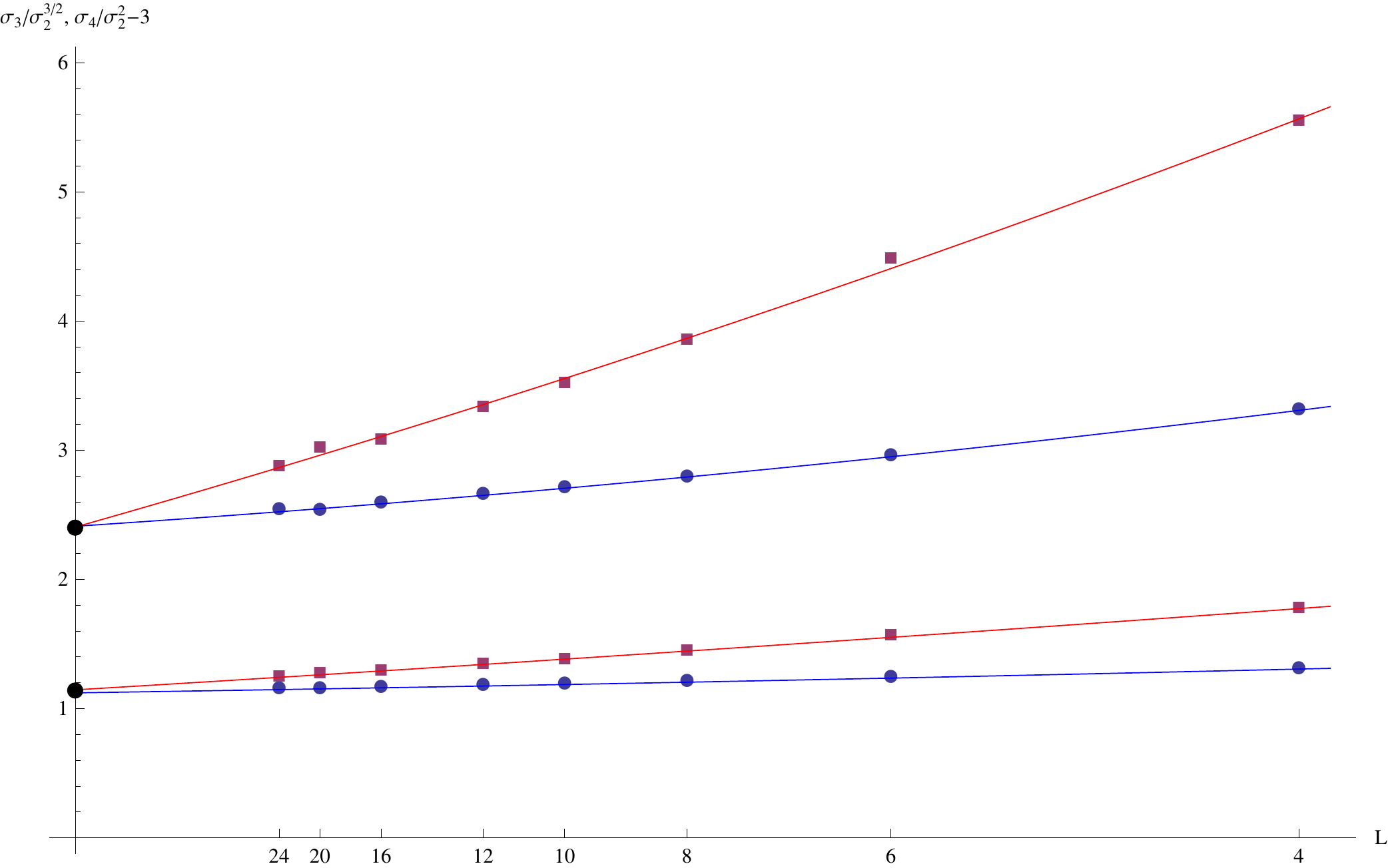}
    \end{center}
  \end{minipage}%
  \begin{minipage}{0.49\textwidth}
    \begin{center}
      \includegraphics[width=0.99\textwidth]{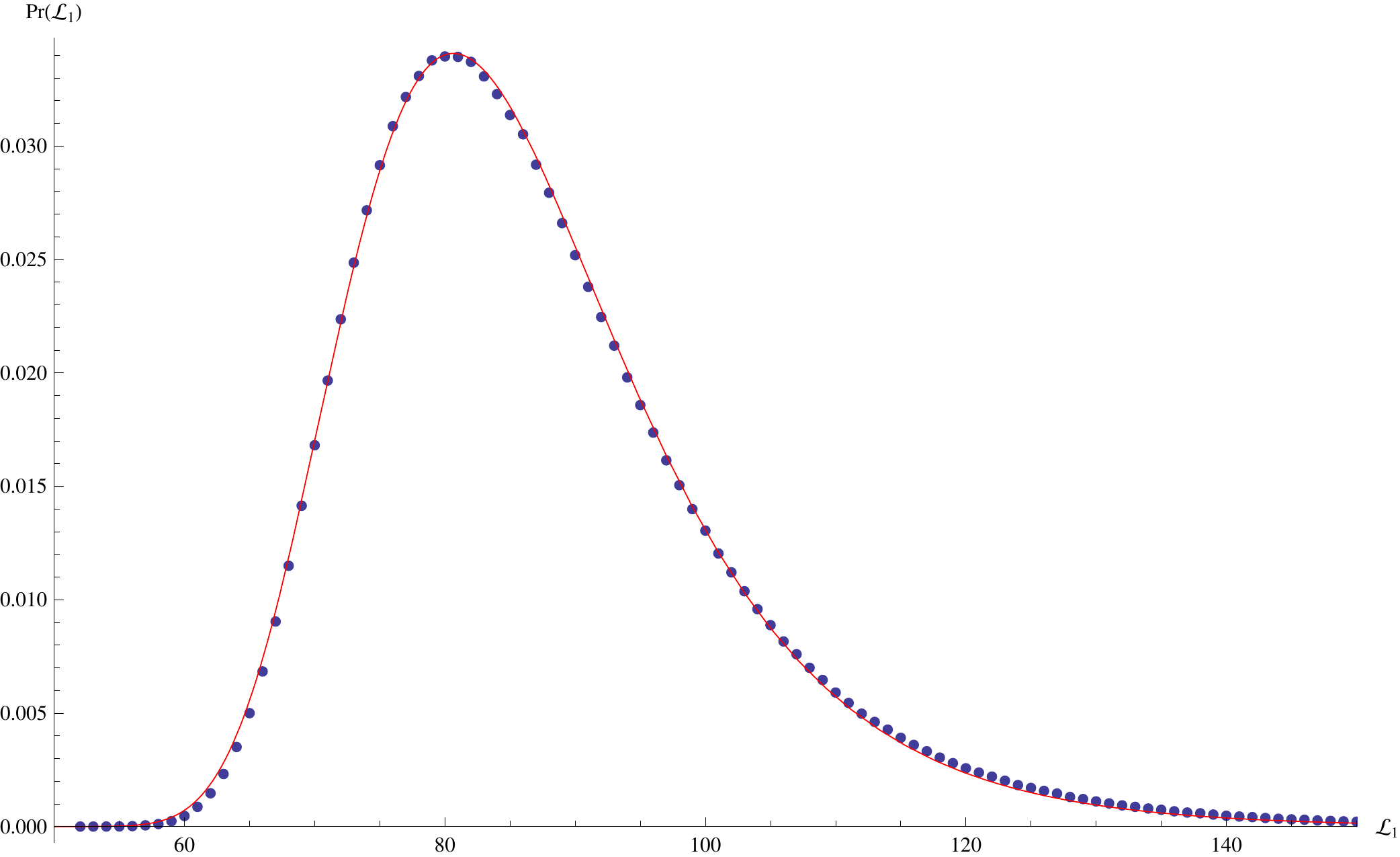}
    \end{center}
  \end{minipage}
  \caption{(Colour online) Left: skewness $\sigma_3/\sigma_2^{3/2}$,
    pointing to $1.1395$, and excess kurtosis $\sigma_4/\sigma_2^2-3$,
    pointing to $2.4$, of the distribution of $\la$ at $K=0.03$ (blue
    circles) and $K=0.08$ (red squares) plotted versus $1/L$ for
    $L=4$, $6$, $8$, $10$, $12$, $16$, $20$, $24$. The fitted curves
    are second degree polynomials. The black points at $x=0$ are the
    extreme value distribution values. Right: Distribution of $\la$
    for $L=16$ at $K=0.08$ (points) and a fitted density function of
    Equation~\eqref{evd} with parameters $a=80.66$ and $b=10.80$.}
  \label{fig:laskhi}
\end{figure}

In \cite{luczak:06} it was shown that $\la$ is distributed as an
extreme-value distribution \cite{de2006extreme}. These have a density
function of the form
\begin{equation}\label{evd}
  f(x) = \exp\left(\frac{a-x}{b} - \exp\left(\frac{a-x}{b}\right)\right)
\end{equation}
given some parameters $a,b$. Extreme value distributions have skewness
$12\sqrt{6}\zeta(3)/\pi^2 = 1.139547\ldots$ and kurtosis $27/5 =
5.4$. In Figure~\ref{fig:laskhi} we show how the skewness and excess
kurtosis scales with $L$ and how they approach these values for two
different temperatures. The right plot of Figure~\ref{fig:laskhi}
shows the distribution of $\la$ at $K=0.08$ for $L=16$. In short, we
have good reason to think that the complete graph behaviour also holds
for 5d in this case.

\section{Negative scaled temperature $\lambda$}

\begin{figure}[!hbt]
  \begin{minipage}{0.49\textwidth}
    \begin{center}
      \includegraphics[width=0.99\textwidth]{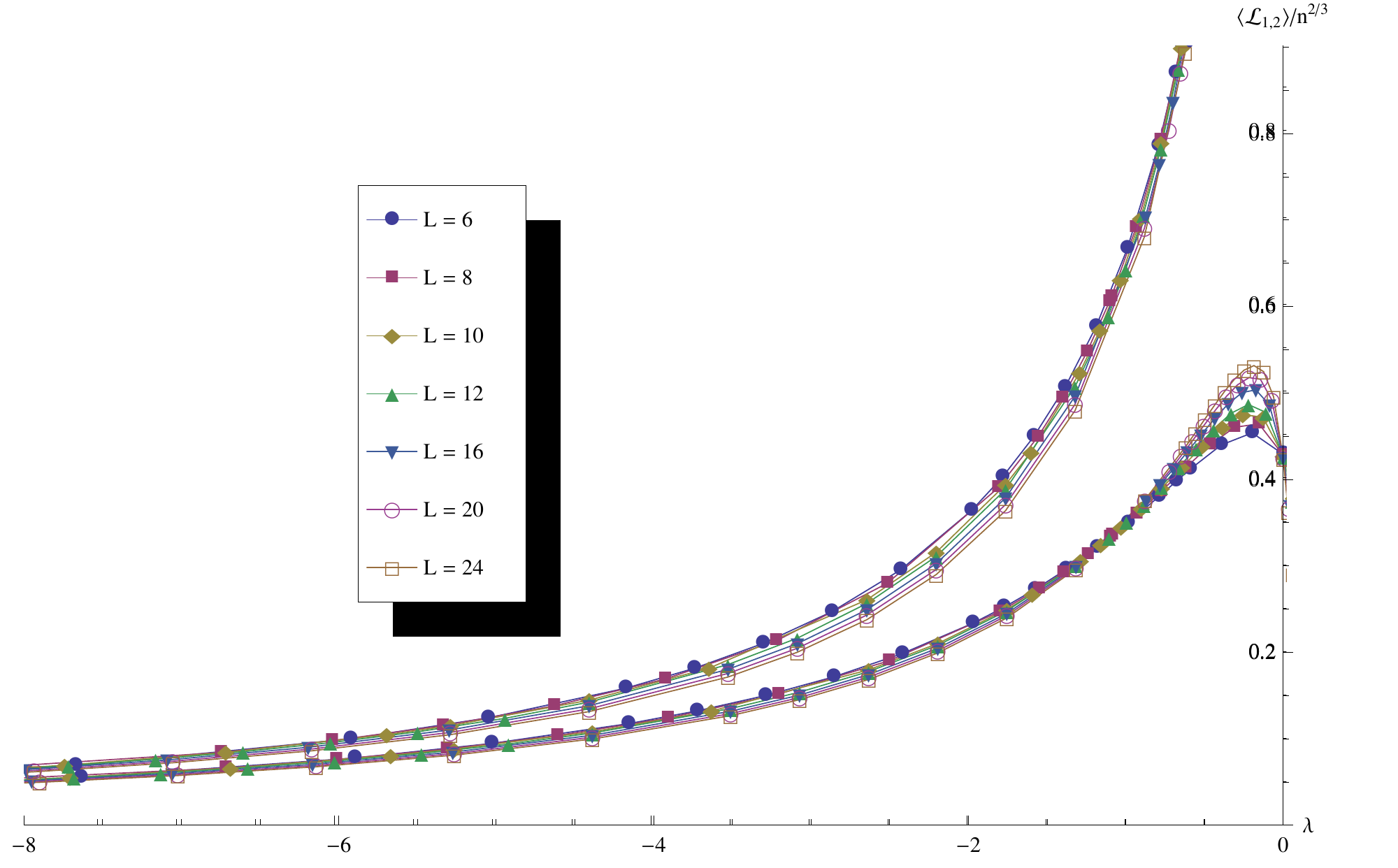}
    \end{center}
  \end{minipage}%
  \begin{minipage}{0.49\textwidth}
    \begin{center}
      \includegraphics[width=0.99\textwidth]{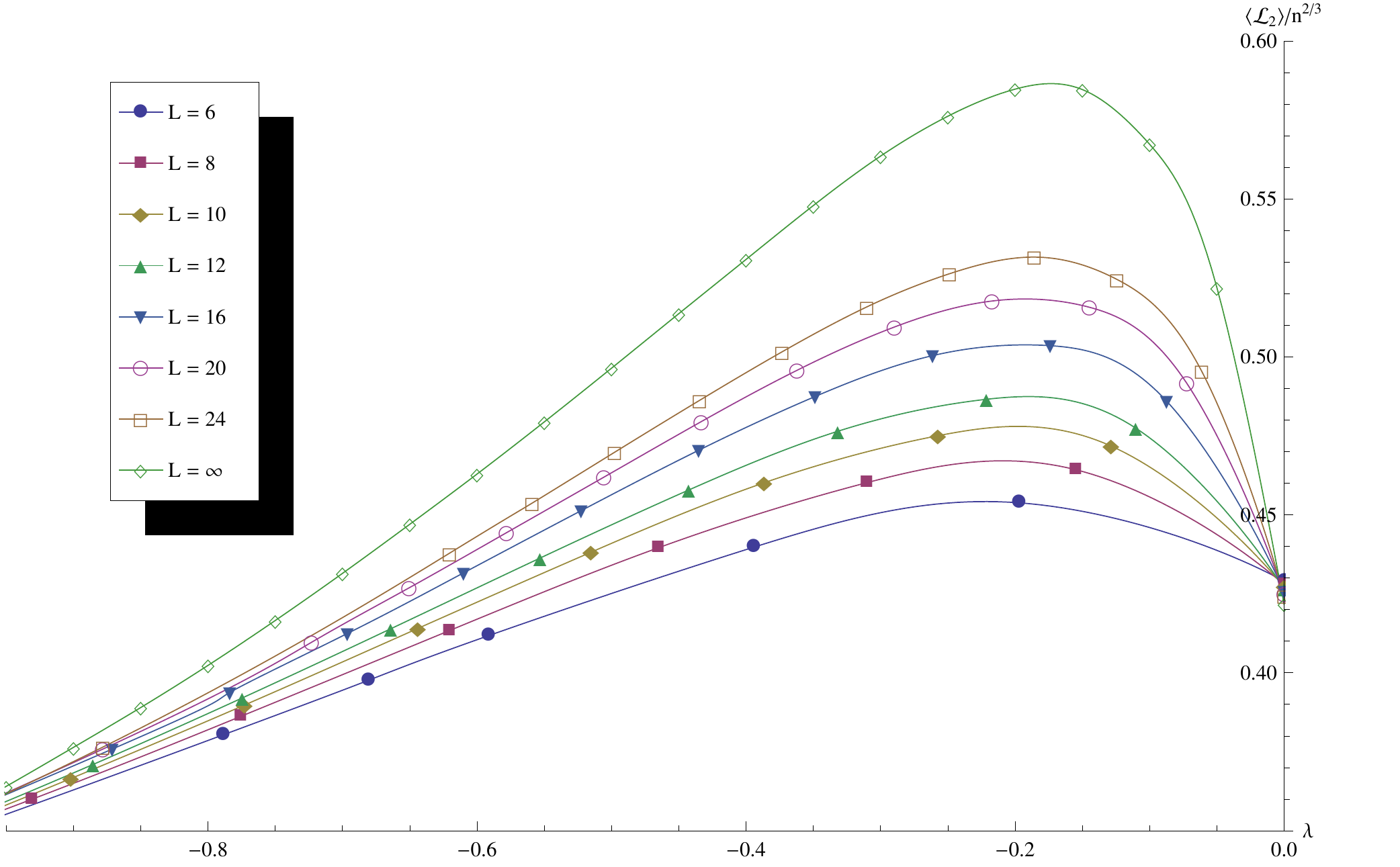}
    \end{center}
  \end{minipage}
  \caption{(Colour online) Left: $\mean\la/n^{2/3}$ (above) and
    $\mean\lb/n^{2/3}$ (below) versus $\lambda$ for $L=6$, $8$, $10$,
    $12$, $16$, $20$, $24$.  Right: $\mean\lb/n^{2/3}$ for $L=6$, $8$,
    $10$, $12$, $16$, $20$, $24$ and $\infty$ (see text).}
  \label{fig:lablambda}
\end{figure}

Here we consider the case of a fixed negative scaled temperature,
$n^{1/3} (K-K_c)/K_c = \lambda < 0$. This falls under the complete
graph case $\varepsilon n^{1/3}\to a < 0$.  In ~\cite{luczak:06} it
was shown that for this particular case the two largest clusters
behave almost surely as
\begin{equation}
  \frac{n^{2/3}}{\omega(n)} \le \lb \le \la \le n^{2/3} \omega(n)
\end{equation}
for all functions $\omega(n)$ tending to infinity. Since $\omega(n)$
is allowed to grow as slowly as we like, thus boxing in $\la$ and
$\lb$, it is quite possible that in fact $\la, \lb \propto n^{2/3}$,
though with different prefactors. Let us test this for the 5D case. In
Figure~\ref{fig:lablambda} we show the normalised mean $\la$ and $\lb$
for different negative $\lambda$. The right plot shows a zoom-in for
$\lb$.

Note here that $\mean\la/n^{2/3}\to\infty$ when $\lambda\to 0^-$
whereas the $\lb$ counterpart actually has a local maximum. The right
panel of Figure~\ref{fig:lablambda} shows a zoom-in for $\lb$, both
the sampled data for $L\le 24$ and also an estimated limit function
based on fitting second degree polynomials to the values at $\lambda$
versus $1/L$. The $\mean\lb/n^{2/3}$ has a distinct local maximum for
all $L\ge 6$ and its location may possibly have zero as limit,
approaching it very slowly. It would take considerably bigger graphs
to settle that question. However, a rough estimate gives that the
limit has the local maximum $0.59$ at $\lambda=-0.17$.  To conclude
this section, we find that the 5D behaviour matches that of the
complete graph case.

\section{Scaled temperature $\kappa$}

The next case is that of $n^{1/2}(K-K_c)/K_c=\kappa$, for constant
$\kappa$. This fits under the complete graph case in ~\cite{luczak:06}
of $\varepsilon n^{1/2}\to c$ for real constants $c$. In this region
the complete graph has different scalings for $\la$ and $\lb$ and we
treat them separately in the following two subsections.

\subsection{The largest cluster}
\begin{figure}[!hbt]
  \begin{minipage}{0.49\textwidth}
    \begin{center}
      \includegraphics[width=0.99\textwidth]{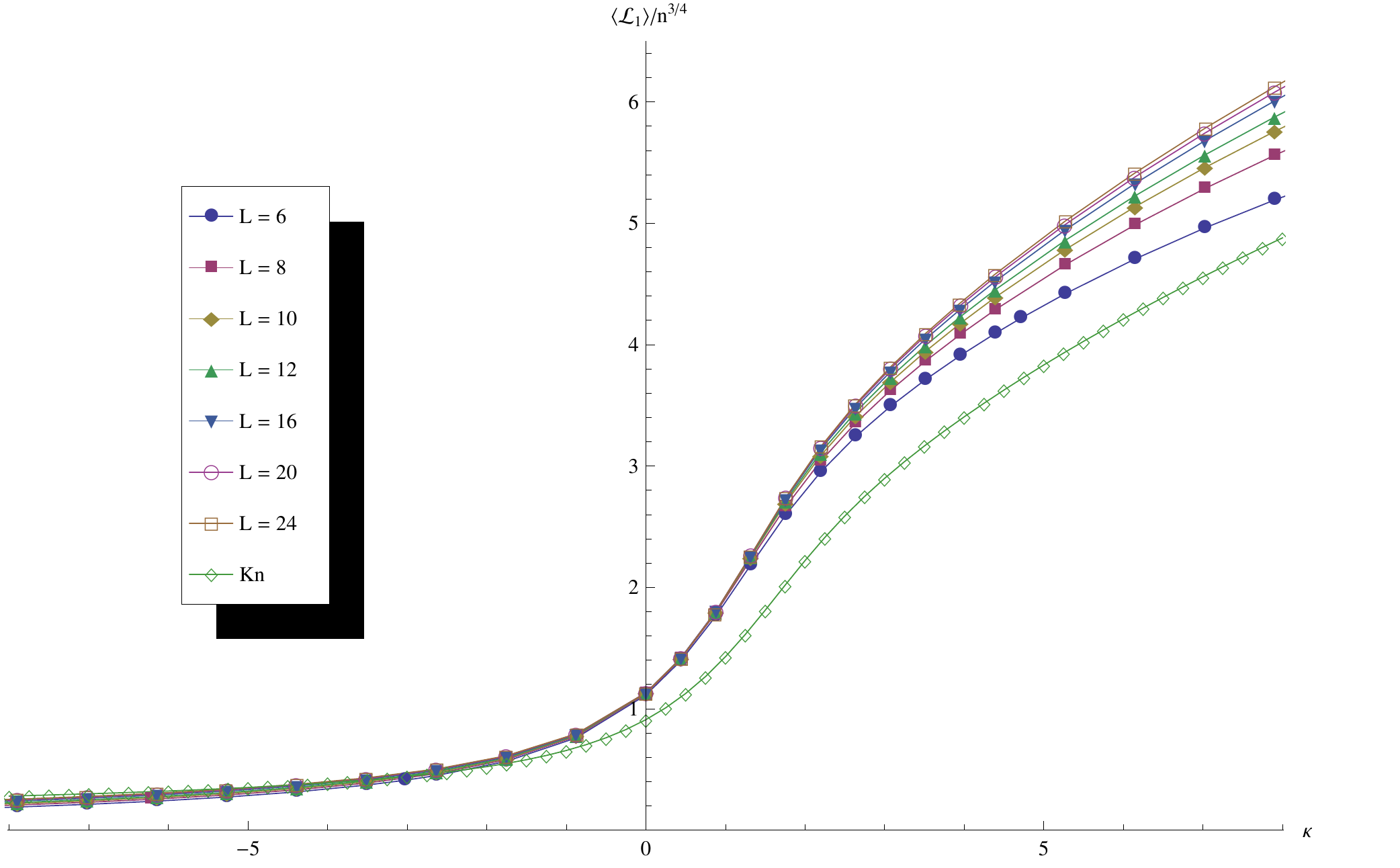}
    \end{center}
  \end{minipage}%
  \begin{minipage}{0.49\textwidth}
    \begin{center}
      \includegraphics[width=0.99\textwidth]{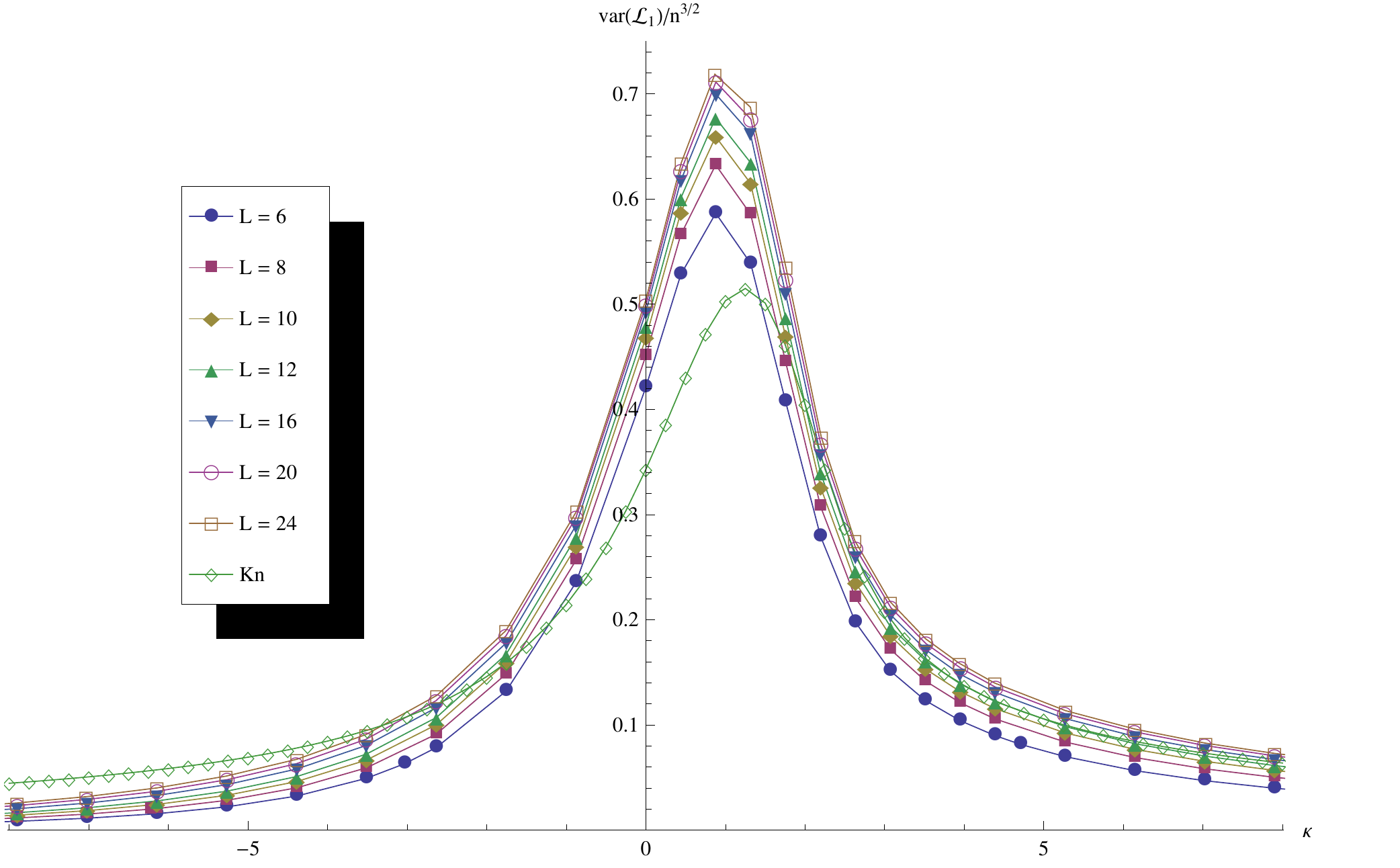}
    \end{center}
  \end{minipage}
  \caption{(Colour online) Left: normalised largest cluster size
    $\mean\la/n^{3/4}$ plotted versus scaled temperature $\kappa$ for
    $L=6$, $8$, $10$, $12$, $16$, $20$, $24$ and the complete graph
    case. Right: normalised variance of the largest cluster size
    $\var\la/n^{3/2}$ plotted versus scaled temperature $\kappa$ for
    $L=6$, $8$, $10$, $12$, $16$, $20$, $24$ and the complete graph
    case.}
  \label{fig:lakappa1}
\end{figure}

In the complete graph case it was shown~\cite{luczak:06} that
\begin{equation}\label{ladist}
  \lim_{n\to\infty} n^{3/4} \Pr(\la=\lfloor a n^{3/4}\rfloor) = 
  \frac{
    \exp(-a^4/12+a^2 c/2)
  }{
    \int_0^{\infty} \exp(-x^4/12+x^2 c/2) \dd x
  }
\end{equation}
and in fact $a$ can be replaced by $a(n)$ with some positive limit
value $a$.  Note that the $\la$-distribution essentially has the same
form as the Ising magnetisation distribution of a complete graph,
which is $\exp(-4a^4/3+2\kappa a^2)$, see \cite{pqpaper}, except that
$\la$ has support only for $a\ge 0$.  This follows since the
spin-state coming from an RC-state is obtained by assigning a random
spin value to each cluster in the RC-state.  Above all this means that
$\mean\la\sim n^{3/4}$ for constant $\kappa$. In
Figure~\ref{fig:lakappa1} we show the normalised mean
$\mean\la/n^{3/4}$ and the normalised variance $\var\la/n^{3/2}$ over
the interval $-8\le\kappa\le 8$. The plot also shows the complete
graph values computed from Equation~\eqref{ladist} when setting
$c=\kappa$. Note how close the complete graph values are to their 5D
counterparts.

The limit of the mean $\la$ is easily found by plotting them for some
$\kappa$ versus $1/L$ and then fitting a polynomial to the points.
There is some very mild correction to scaling at work for $\kappa\ge
0$ but it is easily captured by a second degree polynomial. A slightly
more careful analysis on the case $\kappa=0$, i.e. at $K_c$, based on
fitting second degree polynomials to all 4-subsets of the data points
for $L\ge 6$, results in a median value $\mean\la/n^{3/4}\to
1.1266(6)$, and the error corresponds to the interquartile range.  The
complete graph value at $\kappa=0$ is $0.9098\ldots$.

\begin{figure}[!hbt]
  \begin{minipage}{0.49\textwidth}
    \begin{center}
      \includegraphics[width=0.99\textwidth]{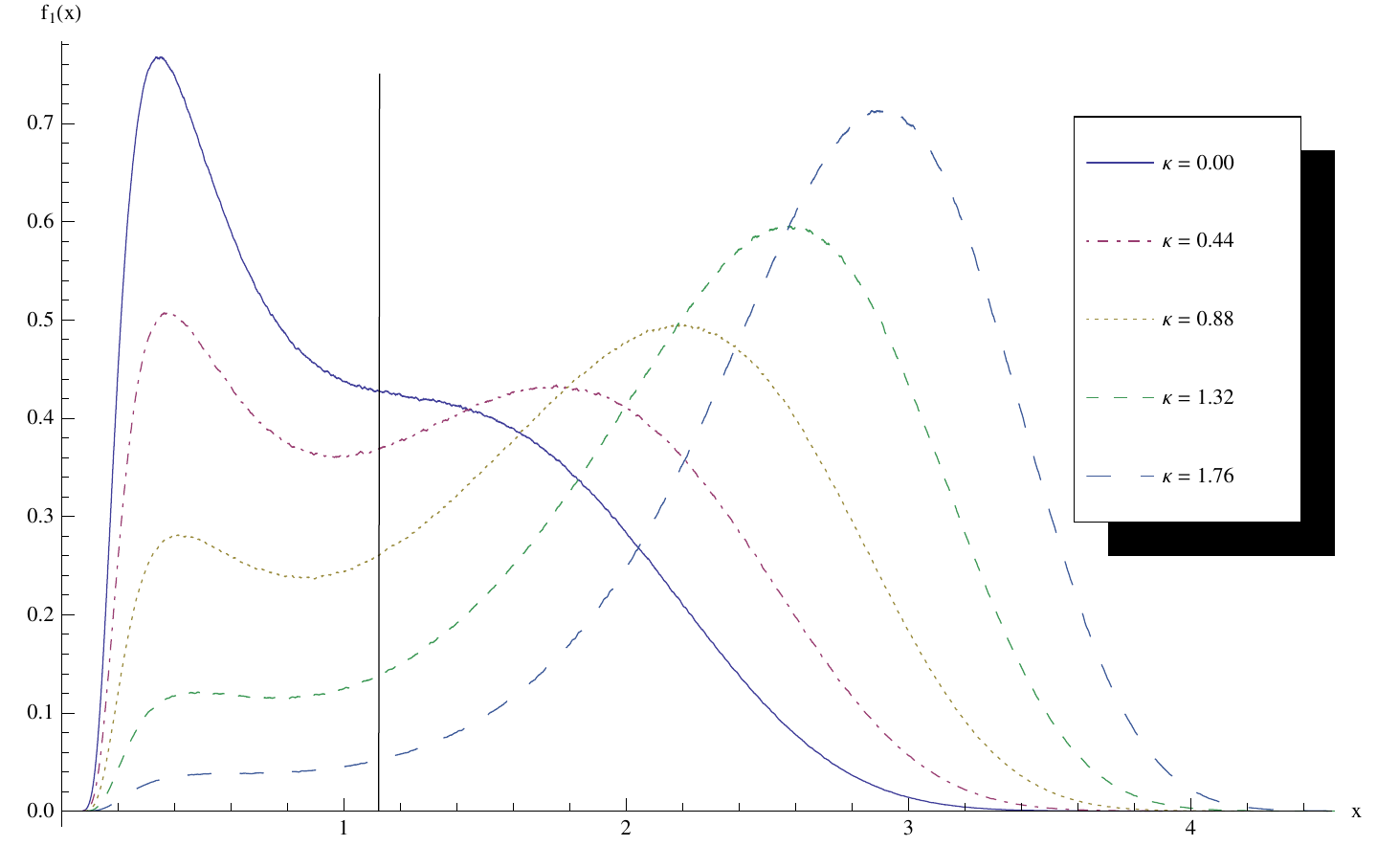}
    \end{center}
  \end{minipage}%
  \begin{minipage}{0.49\textwidth}
    \begin{center}
      \includegraphics[width=0.99\textwidth]{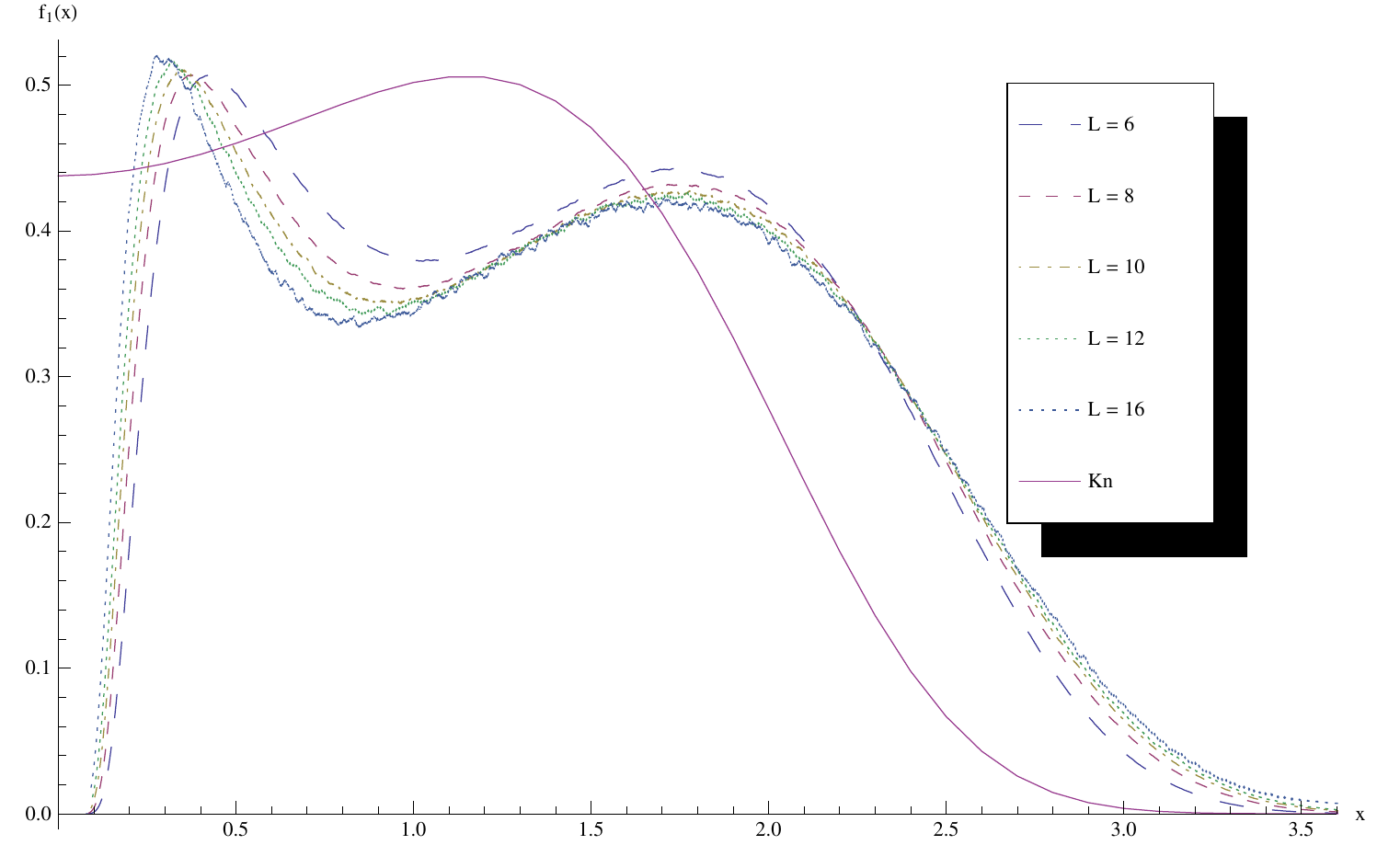}
    \end{center}
  \end{minipage}
  \caption{(Colour online) Left: normalised $\la$-distribution
    $n^{3/4} \Pr(\la)$ versus $\la/n^{3/4}$ for $L=8$ at $\kappa=0$,
    $0.44$, $0.88$, $1.32$ and $1.76$. The black vertical line is
    located at $x=1.126$, the asymptotic mean for $\kappa=0$.  Right:
    normalised $\la$-distribution $n^{3/4} \Pr(\la)$ versus
    $\la/n^{3/4}$ for $L=6$, $8$, $10$, $12$, $16$ and complete graph
    case at $\kappa=0.44$.}
  \label{fig:ladist}
\end{figure}

What about the distribution of $\la$? Figure~\ref{fig:ladist} shows a
normalised form of the $\la$-distribution, $f_1(x)=n^{3/4}
\Pr(\la=\ell)$ where $x=\ell/n^{3/4}$, for a few values of $\kappa$
with $L=8$. Between roughly $0<\kappa<1.8$ the distribution actually
goes through a bimodal phase and this property is not found in the
complete graph case of Equation~\eqref{ladist}. However, it agrees
with the finding in \cite{boundarypaper} that for the Ising model with
cyclic boundary in 5d the energy distribution at the critical
temperature becomes bimodal. The second plot in
Figure~\ref{fig:ladist} shows the distribution at $\kappa=0.44$ for
$L=6,8,10,12,16$ and it clearly shows the distribution retaining its
bimodal form with increasing $L$.

\begin{figure}[!hbt]
  \begin{minipage}{0.49\textwidth}
    \begin{center}
      \includegraphics[width=0.99\textwidth]{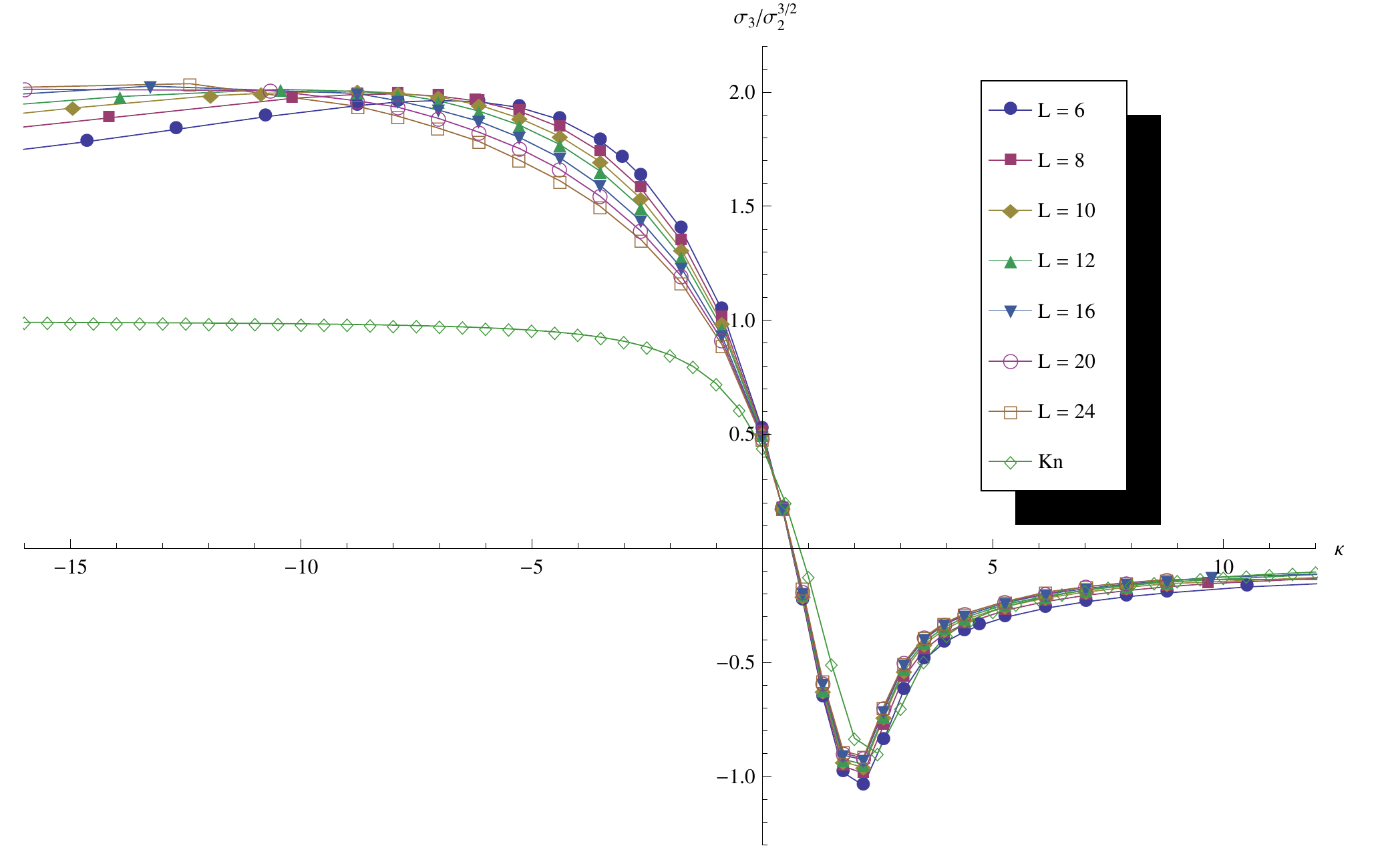}
    \end{center}
  \end{minipage}%
  \begin{minipage}{0.49\textwidth}
    \begin{center}
      \includegraphics[width=0.99\textwidth]{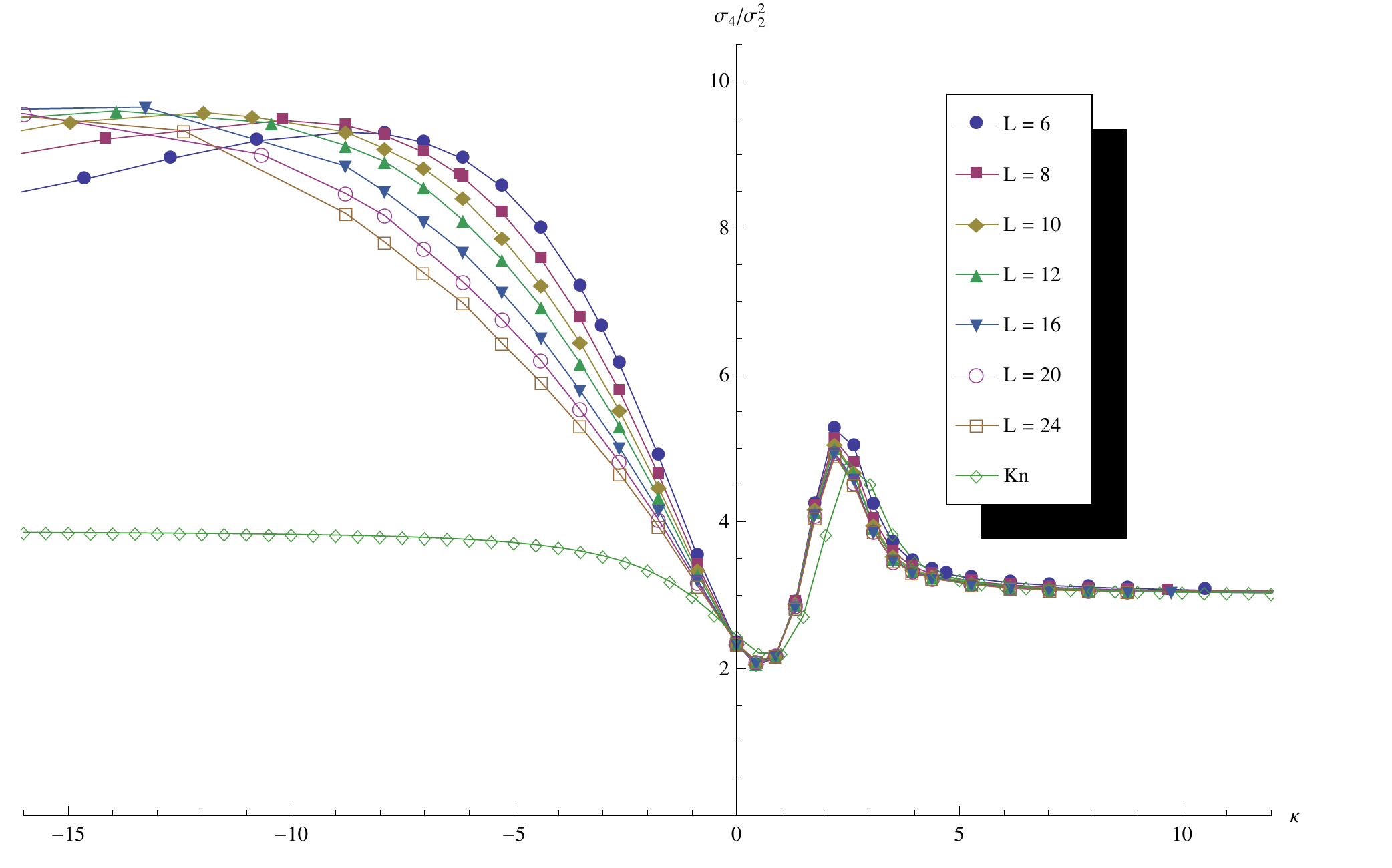}
    \end{center}
  \end{minipage}
  \caption{(Colour online) Left: skewness $\sigma_3/\sigma_2^{3/2}$ of
    the $\la$-distribution plotted versus scaled temperature $\kappa$
    for $L=6$, $8$, $10$, $12$, $16$, $20$, $24$ and omplete graph
    case. Right: kurtosis $\sigma_4/\sigma_2^2$ of the
    $\la$-distribution plotted versus scaled temperature $\kappa$ for
    $L=6$, $8$, $10$, $12$, $16$, $20$, $24$ and complete graph case.}
  \label{fig:lakappa2}
\end{figure}

The skewness and kurtosis of the $\la$-distribution are shown in
Figure~\ref{fig:lakappa2} together with the complete graph
case. Clearly there are distinct limits for each $\kappa$,
occasionally with some mild corrections to scaling, easily captured by
a second degree polynomial.  Perhaps unexpectedly, the data suggest a
limit value of the skewness of about $-2.1$ as $\kappa\to
-\infty$. There is no conflict in the existence of such a limit and
our earlier claim of the skewness taking the extreme value
distribution value of $1.1395\ldots$ in the high-temperature case. It
is simply a matter of taking limits in the right order,
i.e. $\lim_{\kappa\to -\infty} \lim_{L\to\infty}
\sigma_3/\sigma_2^{3/2} \approx 2.1$ and the kurtosis limit is
approximately $10$.  This is in fact also the behaviour of the
complete graph case though it takes different values. As
$\kappa\to-\infty$ its skewness approaches $0.9952\ldots$ and the
kurtosis has the limit $3.869\ldots$.  The two models agree well on
the case $\kappa=0$ though. Here the 5D case has skewness $0.458(6)$
and kurtosis $2.365(4)$ while the complete graph values are
$0.4427\ldots$ and $2.4446\ldots$ respectively.

\subsection{The second largest cluster}

In the complete graph case it was shown~\cite{luczak:06} that
\begin{equation}\label{lbdist}
  \lim_{n\to\infty} \Pr\left(\lb\le \frac{\sqrt{n}\log n}{2 a^2}\right) = 
  \frac{
    \int_a^{\infty}\exp(-x^4/12+x^2 c/2) \dd x
  }{
    \int_0^{\infty} \exp(-x^4/12+x^2 c/2) \dd x
  }
\end{equation}
for $\varepsilon n^{1/2}\to c$. Note that $a$ may be replaced by
function $a(n)$ having some limit $a$. This gives a density function

\begin{equation}\label{lbdist2}
  \lim_{n\to\infty} \sqrt n\log n \Pr\left(\lb=\lfloor b \sqrt n\log n\rfloor\right) = 
  \frac{
    \exp\left(\frac{-1}{48 b^2} + \frac{c}{4 b}\right)
  }{
    2 \sqrt 2 b^{3/2} \int_0^{\infty} \exp(-x^4/12+x^2 c/2) \dd x
  }
\end{equation}
which implies an infinite expectation value. We will thus instead
consider the median value.  So, when we move close enough to $p_c$,
from $n^{1/3} \varepsilon\to c$ to $\varepsilon n^{1/2}\to c$, the
second largest cluster drops in size from $n^{2/3}$ to $\sqrt{n} \log
n$. This also seems to be the case for the 5D case though the
corrections to scaling are quite significant for $\kappa<0$.

\begin{figure}[!hbt]
  \begin{minipage}{0.49\textwidth}
    \begin{center}
      \includegraphics[width=0.99\textwidth]{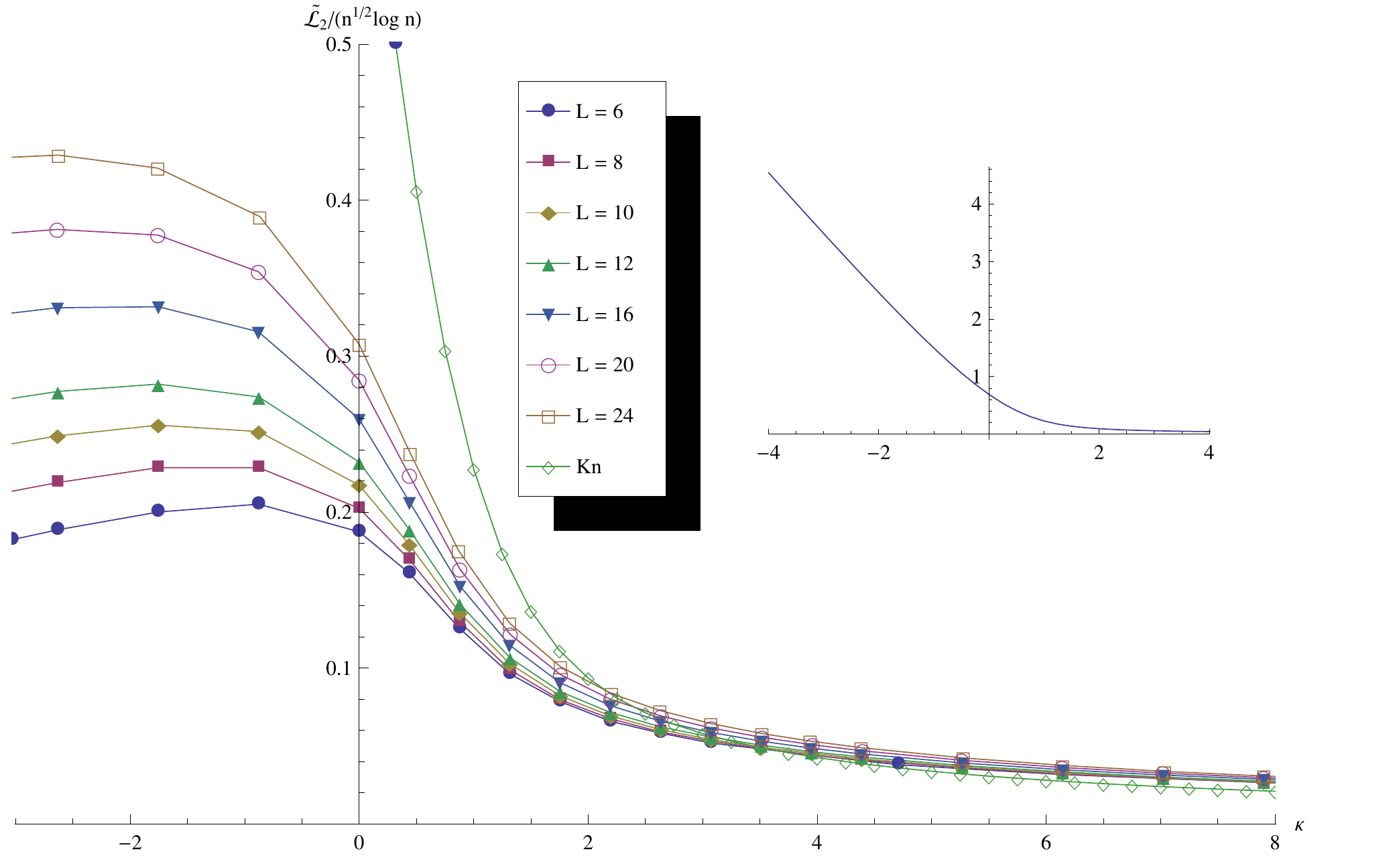}
    \end{center}
  \end{minipage}%
  \begin{minipage}{0.49\textwidth}
    \begin{center}
      \includegraphics[width=0.99\textwidth]{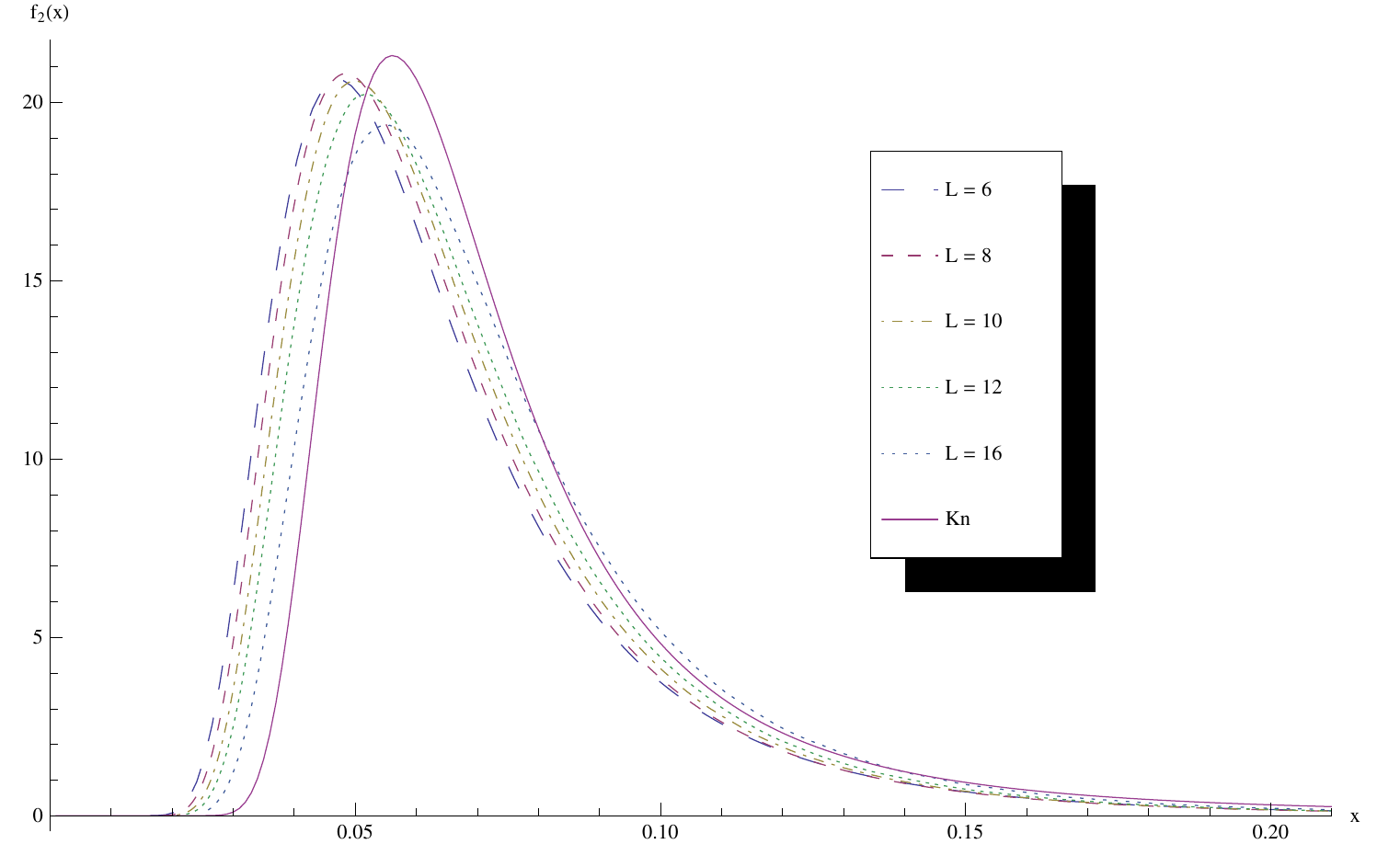}
    \end{center}
  \end{minipage}
  \caption{(Colour online) Left: normalised median second largest
    cluster size $\med\lb/(\sqrt{n} \log n)$ plotted versus scaled
    temperature $\kappa$ for $L=4$, $6$, $8$, $10$, $12$, $16$, $20$,
    $24$ and the complete graph case. The inset shows the complete graph 
    median for a wider range of $\kappa$. Right: normalised
    distribution $f_2(x)$ (see text) of $\lb$, for $L=6$, $8$, $10$,
    $12$, $16$ and the complete graph case of Equation~\eqref{lbdist2} at
    $\kappa=2.63$.}
  \label{fig:lbkappa1}
\end{figure}

Figure~\ref{fig:lbkappa1} shows the normalised mediam $\med\lb/(\sqrt
n \log n)$ versus $\kappa$ for 5D and the complete graph case which
demonstrates this effect. Note that for $\kappa<0$ it would require
enormous graphs to get anything close to the complete graph case. For
$\kappa>0$, however, the two curves quickly agree on the complete
graph behaviour, if not on the actual value. The right plot of the
figure shows a normalised form of the $\lb$-distribution,
$f_2(x)=\sqrt n\log n \Pr(\lb=\ell)$ with $x=\ell/(\sqrt n\log n)$,
for a range of $L$ together with the complete graph distribution in
Equation~\eqref{lbdist2} at $\kappa=2.6$ where the two cases largely
agree.

\begin{figure}[!hbt]
  \begin{minipage}{0.49\textwidth}
    \begin{center}
      \includegraphics[width=0.99\textwidth]{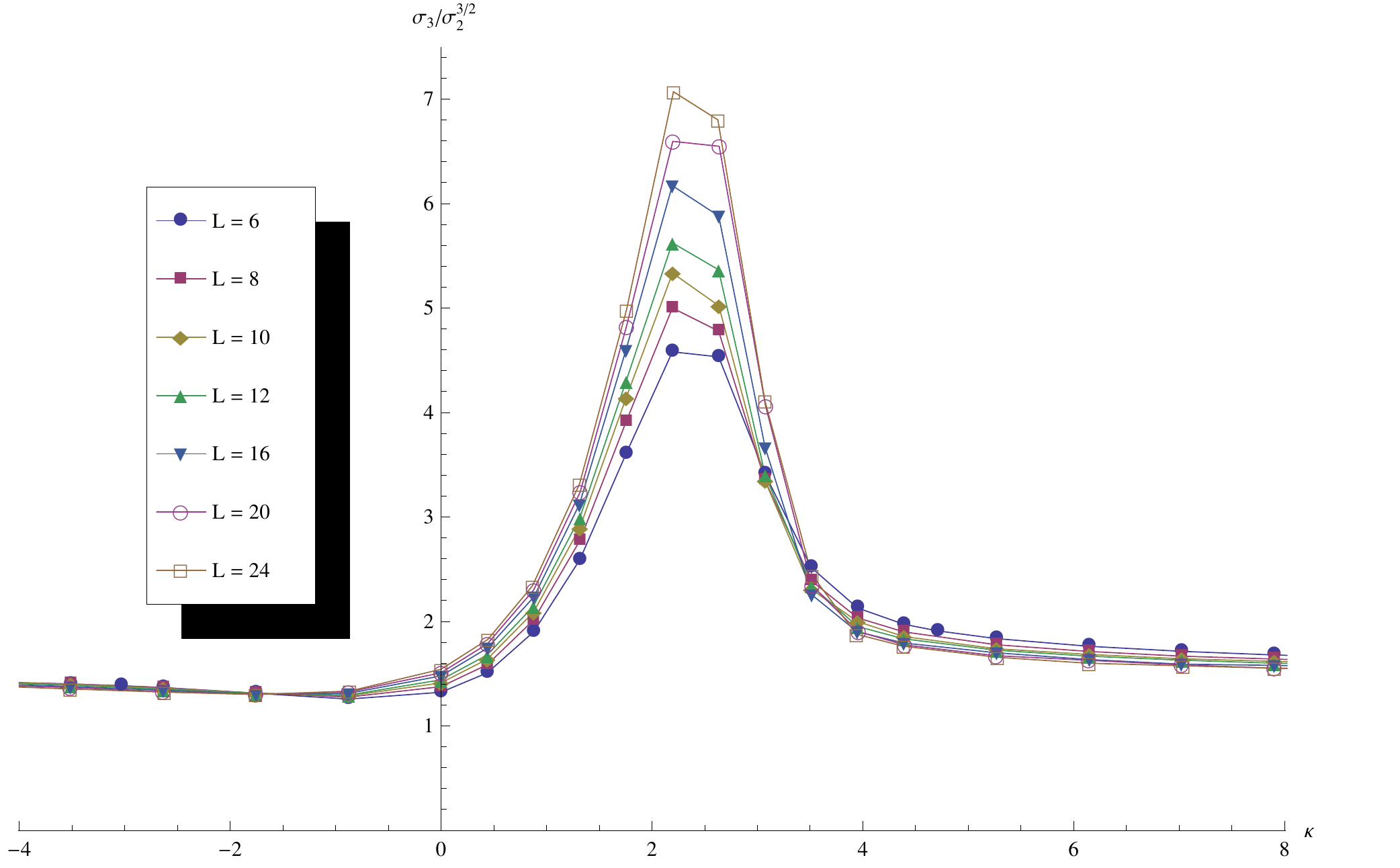}
    \end{center}
  \end{minipage}%
  \begin{minipage}{0.49\textwidth}
    \begin{center}
      \includegraphics[width=0.99\textwidth]{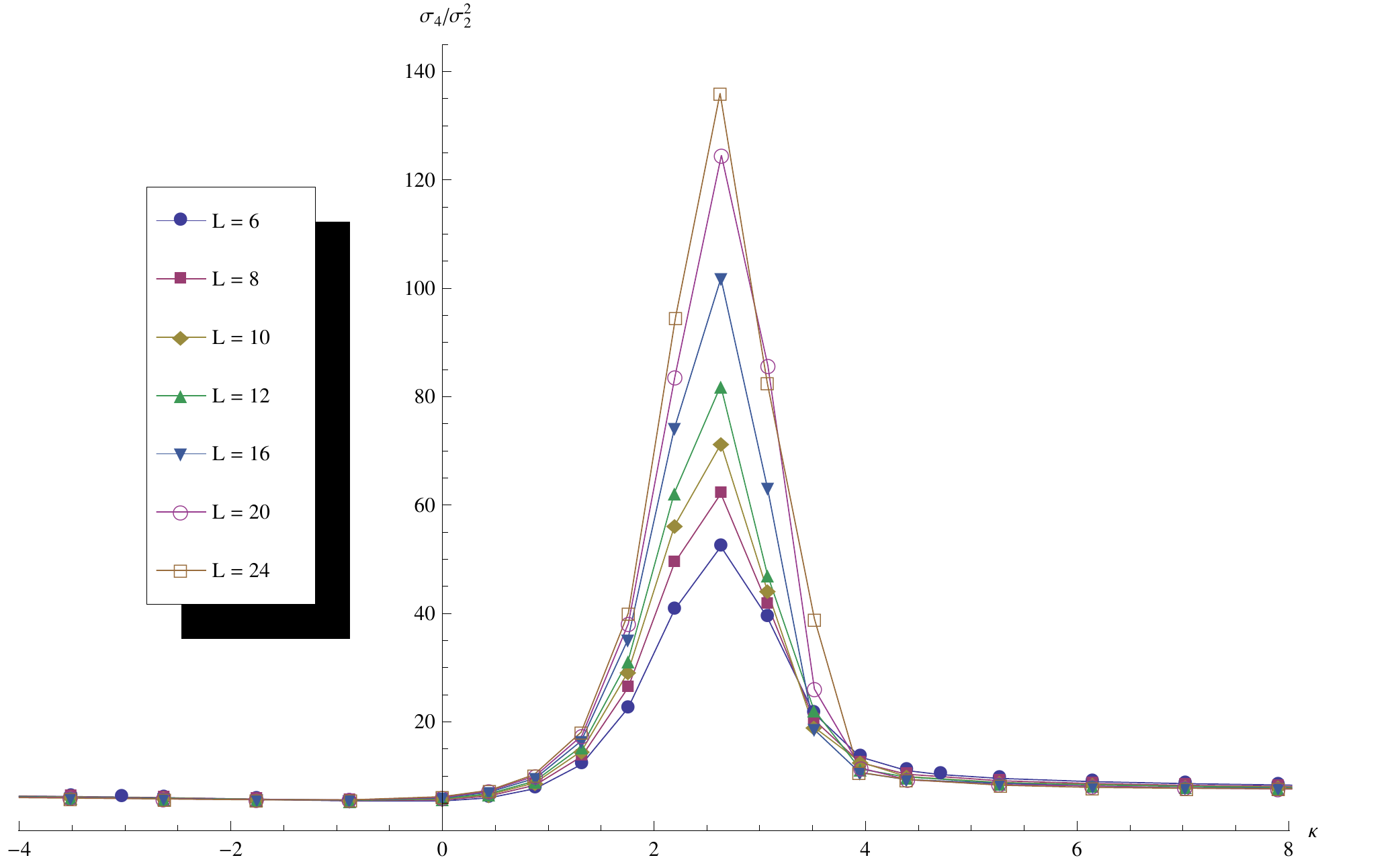}
    \end{center}
  \end{minipage}
  \caption{(Colour online) Left: skewness $\sigma_3/\sigma_2^{3/2}$ of
    the $\lb$-distribution plotted versus $\kappa$ for $L=6$, $8$,
    $10$, $12$, $16$, $20$, $24$.  Right: kurtosis
    $\sigma_4/\sigma_2^2$ of the $\lb$-distribution plotted versus
    $\kappa$ for $L=6$, $8$, $10$, $12$, $16$, $20$, $24$.  }
  \label{fig:lbkappa2}
\end{figure}

If the 5D $\lb$-distribution has a fat tailed distribution like the
complete graph case of Equation~\eqref{lbdist2} this must eventually
show up in some higher moment, and indeed it does. Consider the
skewness and kurtosis of the $\lb$-distribution in
Figure~\ref{fig:lbkappa2}.  Both show a divergent behaviour around
$\kappa\approx 2.5$.  In fact, a very rough estimate suggests that the
peak skewness grows as $2 L^{2/5}$ and the peak kurtosis as
$10L^{4/5}$.

\section{Conclusions}

Our aim has been to compare the scaling of the sizes of the largest
and second largest clusters for the 5D random cluster model with that
seen for the corresponding value of $p$ for the complete graph.  For
the complete graph there are two non-critical regions and inside the
critical window five distinct scaling regions. Due to the limitations
coming from the range of system sizes for which we can simulate the
model we have chosen to work with the high temperature region and two
of the regions from the critical window.

In each of the tested regions we have found that the scaling from the
complete graph agrees well with the observed values from the 5D model,
and in many cases not only the scaling but also the probability
distributions agree well.  The most notable exception is the bimodal
distribution of $\la$ for $\kappa$ close to 0 for the 5D model. This
feature corresponds to the bimodal energy distribution seen for the
Ising model on the same graphs in \cite{boundarypaper}, which also
clearly separates the 5D model with cyclic boundary from the case with
free boundary and the infinite system thermodynamic limit.

We expect this agreement to hold for $d>5$ as well. If we in the
scaling for $\la$ and $\lb$ on the complete graph case replace $n$ by
$V=L^d$ we conjecture that we get the correct scaling for the
$d$-dimensional case with cyclic boundary.

\section{Acknowledgements}
The computations were performed on resources provided by the Swedish
National Infrastructure for Computing (SNIC) at High Performance
Computing Center North (HPC2N) and at Chalmers Centre for
Computational Science and Engineering (C3SE).


\begin{thebibliography}{29}
\expandafter\ifx\csname natexlab\endcsname\relax\def\natexlab#1{#1}\fi
\expandafter\ifx\csname bibnamefont\endcsname\relax
  \def\bibnamefont#1{#1}\fi
\expandafter\ifx\csname bibfnamefont\endcsname\relax
  \def\bibfnamefont#1{#1}\fi
\expandafter\ifx\csname citenamefont\endcsname\relax
  \def\citenamefont#1{#1}\fi
\expandafter\ifx\csname url\endcsname\relax
  \def\url#1{\texttt{#1}}\fi
\expandafter\ifx\csname urlprefix\endcsname\relax\def\urlprefix{URL }\fi
\providecommand{\bibinfo}[2]{#2}
\providecommand{\eprint}[2][]{\url{#2}}

\bibitem[{\citenamefont{Smirnov}(2010)}]{SM1}
\bibinfo{author}{\bibfnamefont{S.}~\bibnamefont{Smirnov}},
  \bibinfo{journal}{Ann. of Math. (2)} \textbf{\bibinfo{volume}{172}},
  \bibinfo{pages}{1435} (\bibinfo{year}{2010}).

\bibitem[{\citenamefont{Chelkak and Smirnov}(2012)}]{SM2}
\bibinfo{author}{\bibfnamefont{D.}~\bibnamefont{Chelkak}} \bibnamefont{and}
  \bibinfo{author}{\bibfnamefont{S.}~\bibnamefont{Smirnov}},
  \bibinfo{journal}{Invent. Math.} \textbf{\bibinfo{volume}{189}},
  \bibinfo{pages}{515} (\bibinfo{year}{2012}).

\bibitem[{\citenamefont{Grimmett}(2004)}]{grimmett2004random}
\bibinfo{author}{\bibfnamefont{G.}~\bibnamefont{Grimmett}},
  \emph{\bibinfo{title}{The random-cluster model}}
  (\bibinfo{publisher}{Springer}, \bibinfo{year}{2004}).

\bibitem[{\citenamefont{Binder}(1985)}]{O1}
\bibinfo{author}{\bibfnamefont{K.}~\bibnamefont{Binder}}, \bibinfo{journal}{Z.
  Phys. B} \textbf{\bibinfo{volume}{61}}, \bibinfo{pages}{13}
  (\bibinfo{year}{1985}).

\bibitem[{\citenamefont{Binder et~al.}(1985)\citenamefont{Binder, Nauenberg,
  Privman, and Young}}]{O2}
\bibinfo{author}{\bibfnamefont{K.}~\bibnamefont{Binder}},
  \bibinfo{author}{\bibfnamefont{M.}~\bibnamefont{Nauenberg}},
  \bibinfo{author}{\bibfnamefont{V.}~\bibnamefont{Privman}}, \bibnamefont{and}
  \bibinfo{author}{\bibfnamefont{A.~P.} \bibnamefont{Young}},
  \bibinfo{journal}{Phys. Rev. B} \textbf{\bibinfo{volume}{31}},
  \bibinfo{pages}{1498} (\bibinfo{year}{1985}).

\bibitem[{\citenamefont{Binder}(1992)}]{O3}
\bibinfo{author}{\bibfnamefont{K.}~\bibnamefont{Binder}}, in
  \emph{\bibinfo{booktitle}{Computational Methods in Field Theory}}, edited by
  \bibinfo{editor}{\bibfnamefont{H.}~\bibnamefont{Gausterer}} \bibnamefont{and}
  \bibinfo{editor}{\bibfnamefont{C.}~\bibnamefont{Lang}}
  (\bibinfo{publisher}{Springer Berlin Heidelberg}, \bibinfo{year}{1992}), vol.
  \bibinfo{volume}{409} of \emph{\bibinfo{series}{Lecture Notes in Physics}},
  pp. \bibinfo{pages}{59--125}.

\bibitem[{\citenamefont{Rudnick
  et~al.}(1985{\natexlab{a}})\citenamefont{Rudnick, Guo, and Jasnow}}]{O4}
\bibinfo{author}{\bibfnamefont{J.}~\bibnamefont{Rudnick}},
  \bibinfo{author}{\bibfnamefont{H.}~\bibnamefont{Guo}}, \bibnamefont{and}
  \bibinfo{author}{\bibfnamefont{D.}~\bibnamefont{Jasnow}},
  \bibinfo{journal}{J. Stat. Phys.} \textbf{\bibinfo{volume}{41}},
  \bibinfo{pages}{353} (\bibinfo{year}{1985}{\natexlab{a}}).

\bibitem[{\citenamefont{Rudnick
  et~al.}(1985{\natexlab{b}})\citenamefont{Rudnick, Gaspari, and Privman}}]{O5}
\bibinfo{author}{\bibfnamefont{J.}~\bibnamefont{Rudnick}},
  \bibinfo{author}{\bibfnamefont{G.}~\bibnamefont{Gaspari}}, \bibnamefont{and}
  \bibinfo{author}{\bibfnamefont{V.}~\bibnamefont{Privman}},
  \bibinfo{journal}{Phys. Rev. B} \textbf{\bibinfo{volume}{32}},
  \bibinfo{pages}{7594} (\bibinfo{year}{1985}{\natexlab{b}}).

\bibitem[{\citenamefont{Rickwardt et~al.}(1994)\citenamefont{Rickwardt,
  Nielaba, and Binder}}]{O6}
\bibinfo{author}{\bibfnamefont{C.}~\bibnamefont{Rickwardt}},
  \bibinfo{author}{\bibfnamefont{P.}~\bibnamefont{Nielaba}}, \bibnamefont{and}
  \bibinfo{author}{\bibfnamefont{K.}~\bibnamefont{Binder}},
  \bibinfo{journal}{Ann. Physik} \textbf{\bibinfo{volume}{506}},
  \bibinfo{pages}{483} (\bibinfo{year}{1994}).

\bibitem[{\citenamefont{Mon}(1996)}]{O7}
\bibinfo{author}{\bibfnamefont{K.~K.} \bibnamefont{Mon}},
  \bibinfo{journal}{Europhys. Lett.} \textbf{\bibinfo{volume}{34}},
  \bibinfo{pages}{399} (\bibinfo{year}{1996}).

\bibitem[{\citenamefont{Parisi and Ruiz-Lorenzo}(1996)}]{O8}
\bibinfo{author}{\bibfnamefont{G.}~\bibnamefont{Parisi}} \bibnamefont{and}
  \bibinfo{author}{\bibfnamefont{J.~J.} \bibnamefont{Ruiz-Lorenzo}},
  \bibinfo{journal}{Phys. Rev. B} \textbf{\bibinfo{volume}{54}},
  \bibinfo{pages}{R3698} (\bibinfo{year}{1996}).

\bibitem[{\citenamefont{Luijten and Bl\"ote}(1996)}]{O9}
\bibinfo{author}{\bibfnamefont{E.}~\bibnamefont{Luijten}} \bibnamefont{and}
  \bibinfo{author}{\bibfnamefont{H.~W.~J.} \bibnamefont{Bl\"ote}},
  \bibinfo{journal}{Phys. Rev. Lett.} \textbf{\bibinfo{volume}{76}},
  \bibinfo{pages}{1557} (\bibinfo{year}{1996}).

\bibitem[{\citenamefont{Bl\"ote and Luijten}(1997)}]{O10}
\bibinfo{author}{\bibfnamefont{H.~W.~J.} \bibnamefont{Bl\"ote}}
  \bibnamefont{and} \bibinfo{author}{\bibfnamefont{E.}~\bibnamefont{Luijten}},
  \bibinfo{journal}{Europhys. Lett.} \textbf{\bibinfo{volume}{38}},
  \bibinfo{pages}{565} (\bibinfo{year}{1997}).

\bibitem[{\citenamefont{Luijten et~al.}(1999)\citenamefont{Luijten, Binder, and
  Bl\"ote}}]{O11}
\bibinfo{author}{\bibfnamefont{E.}~\bibnamefont{Luijten}},
  \bibinfo{author}{\bibfnamefont{K.}~\bibnamefont{Binder}}, \bibnamefont{and}
  \bibinfo{author}{\bibfnamefont{H.}~\bibnamefont{Bl\"ote}},
  \bibinfo{journal}{Eur. Phys. J. B} \textbf{\bibinfo{volume}{9}},
  \bibinfo{pages}{289} (\bibinfo{year}{1999}).

\bibitem[{\citenamefont{Jones and Young}(2005)}]{O12}
\bibinfo{author}{\bibfnamefont{J.~L.} \bibnamefont{Jones}} \bibnamefont{and}
  \bibinfo{author}{\bibfnamefont{A.~P.} \bibnamefont{Young}},
  \bibinfo{journal}{Phys. Rev. B} \textbf{\bibinfo{volume}{71}},
  \bibinfo{pages}{174438} (\bibinfo{year}{2005}).

\bibitem[{\citenamefont{Lundow and Markstr\"om}(2011)}]{boundarypaper}
\bibinfo{author}{\bibfnamefont{P.~H.} \bibnamefont{Lundow}} \bibnamefont{and}
  \bibinfo{author}{\bibfnamefont{K.}~\bibnamefont{Markstr\"om}},
  \bibinfo{journal}{Nucl. Phys. B} \textbf{\bibinfo{volume}{845}},
  \bibinfo{pages}{120 } (\bibinfo{year}{2011}).

\bibitem[{\citenamefont{Berche et~al.}(2008)\citenamefont{Berche, Chatelain,
  Dhall, Kenna, Low, and Walter}}]{berche:08}
\bibinfo{author}{\bibfnamefont{B.}~\bibnamefont{Berche}},
  \bibinfo{author}{\bibfnamefont{C.}~\bibnamefont{Chatelain}},
  \bibinfo{author}{\bibfnamefont{C.}~\bibnamefont{Dhall}},
  \bibinfo{author}{\bibfnamefont{R.}~\bibnamefont{Kenna}},
  \bibinfo{author}{\bibfnamefont{R.}~\bibnamefont{Low}}, \bibnamefont{and}
  \bibinfo{author}{\bibfnamefont{J.-C.} \bibnamefont{Walter}},
  \bibinfo{journal}{J. Stat. Mech.} \textbf{\bibinfo{volume}{2008}},
  \bibinfo{pages}{P11010} (\bibinfo{year}{2008}).

\bibitem[{\citenamefont{Aizenman}(1997)}]{aizenman:97}
\bibinfo{author}{\bibfnamefont{M.}~\bibnamefont{Aizenman}},
  \bibinfo{journal}{Nucl. Phys. B} \textbf{\bibinfo{volume}{485}},
  \bibinfo{pages}{551} (\bibinfo{year}{1997}).

\bibitem[{\citenamefont{Heydenreich and van~der Hofstad}(2007)}]{HH:07}
\bibinfo{author}{\bibfnamefont{M.}~\bibnamefont{Heydenreich}} \bibnamefont{and}
  \bibinfo{author}{\bibfnamefont{R.}~\bibnamefont{van~der Hofstad}},
  \bibinfo{journal}{Comm. Math. Phys.} \textbf{\bibinfo{volume}{270}},
  \bibinfo{pages}{335} (\bibinfo{year}{2007}).

\bibitem[{\citenamefont{Heydenreich and van~der Hofstad}(2011)}]{HH:11}
\bibinfo{author}{\bibfnamefont{M.}~\bibnamefont{Heydenreich}} \bibnamefont{and}
  \bibinfo{author}{\bibfnamefont{R.}~\bibnamefont{van~der Hofstad}},
  \bibinfo{journal}{Probab. Theory Related Fields}
  \textbf{\bibinfo{volume}{149}}, \bibinfo{pages}{397} (\bibinfo{year}{2011}).

\bibitem[{\citenamefont{Borgs et~al.}(2005{\natexlab{a}})\citenamefont{Borgs,
  Chayes, van~der Hofstad, Slade, and Spencer}}]{MR2155704}
\bibinfo{author}{\bibfnamefont{C.}~\bibnamefont{Borgs}},
  \bibinfo{author}{\bibfnamefont{J.~T.} \bibnamefont{Chayes}},
  \bibinfo{author}{\bibfnamefont{R.}~\bibnamefont{van~der Hofstad}},
  \bibinfo{author}{\bibfnamefont{G.}~\bibnamefont{Slade}}, \bibnamefont{and}
  \bibinfo{author}{\bibfnamefont{J.}~\bibnamefont{Spencer}},
  \bibinfo{journal}{Random Struct. Algorithms} \textbf{\bibinfo{volume}{27}},
  \bibinfo{pages}{137} (\bibinfo{year}{2005}{\natexlab{a}}).

\bibitem[{\citenamefont{Borgs et~al.}(2005{\natexlab{b}})\citenamefont{Borgs,
  Chayes, van~der Hofstad, Slade, and Spencer}}]{MR2165583}
\bibinfo{author}{\bibfnamefont{C.}~\bibnamefont{Borgs}},
  \bibinfo{author}{\bibfnamefont{J.~T.} \bibnamefont{Chayes}},
  \bibinfo{author}{\bibfnamefont{R.}~\bibnamefont{van~der Hofstad}},
  \bibinfo{author}{\bibfnamefont{G.}~\bibnamefont{Slade}}, \bibnamefont{and}
  \bibinfo{author}{\bibfnamefont{J.}~\bibnamefont{Spencer}},
  \bibinfo{journal}{Ann. Probab.} \textbf{\bibinfo{volume}{33}},
  \bibinfo{pages}{1886} (\bibinfo{year}{2005}{\natexlab{b}}).

\bibitem[{\citenamefont{Borgs et~al.}(2006)\citenamefont{Borgs, Chayes, van~der
  Hofstad, Slade, and Spencer}}]{MR2260845}
\bibinfo{author}{\bibfnamefont{C.}~\bibnamefont{Borgs}},
  \bibinfo{author}{\bibfnamefont{J.~T.} \bibnamefont{Chayes}},
  \bibinfo{author}{\bibfnamefont{R.}~\bibnamefont{van~der Hofstad}},
  \bibinfo{author}{\bibfnamefont{G.}~\bibnamefont{Slade}}, \bibnamefont{and}
  \bibinfo{author}{\bibfnamefont{J.}~\bibnamefont{Spencer}},
  \bibinfo{journal}{Combinatorica} \textbf{\bibinfo{volume}{26}},
  \bibinfo{pages}{395} (\bibinfo{year}{2006}).

\bibitem[{\citenamefont{Bollob{\'a}s et~al.}(1996)\citenamefont{Bollob{\'a}s,
  Grimmett, and Janson}}]{MR1376340}
\bibinfo{author}{\bibfnamefont{B.}~\bibnamefont{Bollob{\'a}s}},
  \bibinfo{author}{\bibfnamefont{G.}~\bibnamefont{Grimmett}}, \bibnamefont{and}
  \bibinfo{author}{\bibfnamefont{S.}~\bibnamefont{Janson}},
  \bibinfo{journal}{Probab. Theory Related Fields}
  \textbf{\bibinfo{volume}{104}}, \bibinfo{pages}{283} (\bibinfo{year}{1996}).

\bibitem[{\citenamefont{Luczak and Luczak}(2006)}]{luczak:06}
\bibinfo{author}{\bibfnamefont{M.~J.} \bibnamefont{Luczak}} \bibnamefont{and}
  \bibinfo{author}{\bibfnamefont{T.}~\bibnamefont{Luczak}},
  \bibinfo{journal}{Random Struct. Algorithms} \textbf{\bibinfo{volume}{28}},
  \bibinfo{pages}{215} (\bibinfo{year}{2006}).

\bibitem[{\citenamefont{Newman and Barkema}(1999)}]{newman-barkema}
\bibinfo{author}{\bibfnamefont{M.~E.~J.} \bibnamefont{Newman}}
  \bibnamefont{and} \bibinfo{author}{\bibfnamefont{G.~T.}
  \bibnamefont{Barkema}}, \emph{\bibinfo{title}{Monte {C}arlo methods in
  statistical physics}} (\bibinfo{publisher}{The Clarendon Press Oxford
  University Press}, \bibinfo{address}{New York}, \bibinfo{year}{1999}), ISBN
  \bibinfo{isbn}{0-19-851796-3; 0-19-851797-1}.

\bibitem[{\citenamefont{Wolff}(1989)}]{wolff:89}
\bibinfo{author}{\bibfnamefont{U.}~\bibnamefont{Wolff}},
  \bibinfo{journal}{Phys. Rev. Lett} \textbf{\bibinfo{volume}{62}},
  \bibinfo{pages}{361} (\bibinfo{year}{1989}).

\bibitem[{\citenamefont{de~Haan and Ferreira}(2006)}]{de2006extreme}
\bibinfo{author}{\bibfnamefont{L.}~\bibnamefont{de~Haan}} \bibnamefont{and}
  \bibinfo{author}{\bibfnamefont{A.}~\bibnamefont{Ferreira}},
  \emph{\bibinfo{title}{Extreme Value Theory: An Introduction}}, Springer
  Series in Operations Research and Financial Engineering
  (\bibinfo{publisher}{Springer}, \bibinfo{year}{2006}).

\bibitem[{\citenamefont{Lundow and Rosengren}(2010)}]{pqpaper}
\bibinfo{author}{\bibfnamefont{P.~H.} \bibnamefont{Lundow}} \bibnamefont{and}
  \bibinfo{author}{\bibfnamefont{A.}~\bibnamefont{Rosengren}},
  \bibinfo{journal}{Phil. Mag.} \textbf{\bibinfo{volume}{90}},
  \bibinfo{pages}{3313} (\bibinfo{year}{2010}).

\end{thebibliography}

\end{document}